\documentclass[sn-mathphys,Numbered]{sn-jnl}
\usepackage{graphicx}%
\usepackage{subcaption}
\usepackage{multirow}%
\usepackage{amsmath,amssymb,amsfonts}%
\usepackage{amsthm}%
\usepackage{mathrsfs}%
\usepackage[title]{appendix}%
\usepackage{xcolor}%
\usepackage{textcomp}%
\usepackage{manyfoot}%
\usepackage{booktabs}%
\usepackage{algorithm}%
\usepackage{algorithmicx}%
\usepackage{algpseudocode}%
\usepackage{listings}%
\usepackage{placeins}
\def\al{\alpha} 
\def\be{\beta}

\def\ep{\epsilon}
\def\ze{\zeta}

\def\la{\lambda}

\def\si{\sigma}

\def\De{\Delta}

\def\pa{\partial}

\newcommand{\ben}{\begin{equation}}
\newcommand{\een}{\end{equation}}
\newcommand{\bea}{\begin{eqnarray}}
\newcommand{\eea}{\end{eqnarray}}
\newcommand{\ba}{\begin{array}}
\newcommand{\ea}{\end{array}}
\newcommand{\bit}{\begin{itemize}}
\newcommand{\eit}{\end{itemize}}

\newcommand{\bp}{\textbf{p}}

\newcommand{\mcL}{\mathcal{L}}

\definecolor{red}{rgb}{1,0,0}

\def\Tr#1{\mbox{Tr}\left(#1\right)}

\newcommand{\Ec}{E_\text{c}} 		

\newcommand{\fAB}{f_\text{AB}}
\newcommand{\fA}{f_\text{A}}
\newcommand{\fB}{f_\text{B}}

\newcommand{\Gi}{\text{Gi}}
\newcommand{\hethree}{$^3$He}

\newcommand{\kb}{k_\text{B}}		
\newcommand{\kB}{k_\text{B}}		

\newcommand{\noise}{\zeta}  
\newcommand{\PPCP}{P_\text{PCP}} 

\newcommand{\Rc}{R_\text{c}}

\newcommand{\siAB}{\si_\text{AB}}

\newcommand{\Tc}{T_\text{c}}		
\newcommand{\TAB}{T_\text{AB}}	
\newcommand{\tGL}{t_\text{GL}}	

\newcommand{\vF}{v_\text{F}}		

\newcommand{\vw}{v_\text{w}}		

\newcommand{\xiGL}{\xi_\text{GL}}		

\newcommand{\bd}{\textbf{d}}
\newcommand{\bm}{\textbf{m}}
\newcommand{\bn}{\textbf{n}}
\newcommand{\bl}{\textbf{l}}
\newcommand{\br}{\textbf{r}}

\begin{document}
\title[A-B transition in superfluid $^3$He
and cosmological phase transitions]{A-B transition in superfluid $^3$He
and cosmological phase transitions}
\author*[1,2]{\fnm{Mark} \sur{Hindmarsh}}
\email{m.b.hindmarsh@sussex.ac.uk}
\author[3]{\fnm{J. A.} \sur{Sauls}}
\author[1,2]{\fnm{Kuang} \sur{Zhang}}
\author[4]{\fnm{S.} \sur{Autti}}
\author[4]{\fnm{Richard P.} \sur{Haley}}
\author[5]{\fnm{Petri J.} \sur{Heikkinen}}
\author[1]{\fnm{Stephan J.} \sur{Huber}}
\author[5]{\fnm{Lev V.} \sur{Levitin}}
\author[6,7]{\fnm{Asier} \sur{Lopez-Eiguren}}
\author[4]{\fnm{Adam J.} \sur{Mayer}}
\author[2]{\fnm{Kari} \sur{Rummukainen}}
\author[5]{\fnm{John} \sur{Saunders}}
\author[4]{\fnm{Dmitry} \sur{Zmeev}}

\affil[1]{
Department of Physics and Astronomy,
University of Sussex, Falmer, Brighton BN1 9QH,
UK}
\affil[2]{
Department of Physics and Helsinki Institute of Physics,
PL 64, 
FI-00014 University of Helsinki,
Finland}
\affil[3]{
Hearne Institute of Theoretical Physics, Department of Physics \& Astronomy,
Louisiana State University, Baton Rouge LA 70803, USA}
\affil[4]{
Department of Physics, Lancaster University, Lancaster LA1 4YB, UK}
\affil[5]{
Department of Physics, Royal Holloway, University of London, Egham TW20 0EX, UK}
\affil[6]{
Department of Physics,
University of the Basque Country UPV/EHU, 48080 Bilbao, Spain}
\affil[7]{
EHU Quantum Center, University of Basque Country, UPV/EHU}

\date{\today}

\abstract{
First order phase transitions in the very early universe are a prediction of many extensions of the Standard Model of particle physics and could provide the departure from equilibrium needed for a dynamical explanation of the baryon asymmetry of the Universe. They could also produce gravitational waves of a frequency observable by future space-based detectors such as the Laser Interferometer Space Antenna (LISA). All calculations of the gravitational wave power spectrum rely on a relativistic version of the classical nucleation theory of Cahn-Hilliard and Langer, due to Coleman and Linde. The high purity and precise control of pressure and temperature achievable in the laboratory made the first-order A to B transition of superfluid $^3$He ideal for test of classical nucleation theory. As Leggett and others have noted the theory fails dramatically. The lifetime of the metastable A phase is measurable, typically of order minutes to hours, far faster than classical nucleation theory predicts. If the nucleation of B phase from the supercooled A phase is due to a new, rapid intrinsic mechanism that would have implications for first-order cosmological phase transitions as well as predictions for gravitational wave (GW) production in the early universe. Here we discuss studies of the A-B phase transition dynamics in $^3$He, both experimental and theoretical, and show how the computational technology for cosmological phase transition can be used to simulate the dynamics of the A\nobreakdash-\hspace{0pt}B transition, support the experimental investigations of the A-B transition in the QUEST-DMC collaboration with the goal of identifying and quantifying the mechanism(s) responsible for nucleation of stable phases in ultra-pure metastable quantum phases.
}

\maketitle

\section{Introduction}

First order phase transitions are predicted to occur during cooling in the early universe as a signal of physics beyond the Standard Model. The transition is expected to proceed by the nucleation of bubbles of the stable phase by quantum or thermal fluctuations after supercoooling. These bubbles grow rapidly due to the pressure difference between inside and outside, and subsequently merge. The ``fizz'' generates pressure waves, which on collision produce shear stresses with non-vanishing quadrupole moment, and hence gravitational waves \cite{Witten:1984rs,Hogan:1986qda} (see \cite{Hindmarsh:2020hop} for a review). 

The power spectrum of the gravitational waves depends on a relatively small number of equilibrium and near-equilibrium properties of the phase transition, at least in the case where the super-cooled transition happens fairly close to the critical temperature. Two of the most important are the temperature at which bubbles nucleate, and the duration of the transition. Both can be computed from the bubble nucleation rate per unit volume as a function of temperature. In the standard approach, the nucleation rate density is computed in homogeneous nucleation theory \cite{cahn1959free,langer1969statistical} adapted to relativistic quantum field theory \cite{Coleman:1977py,Linde:1980tt}.  

Given the importance of cosmological nucleation theory for the prediction of GWs in the early universe, and its close relation to nucleation theory developed for laboratory systems, it is important to test the theory against experiment.  Bubble nucleation observed in nature and in standard laboratory systems usually happens around seeds: small particles in suspension or surface irregularities. Testing nucleation theory in the bulk requires extremely pure systems in containers with very smooth walls. 
Thus, the ideal system for such a test is superfluid \hethree, which 
in zero magnetic field has two phases distinguished by their residual symmetries, and a first order transition between them. 
The transition from a normal Fermi liquid to the superfluid takes place between 1--2.5 mK, at which temperatures superfluid \hethree\ is essentially pure. Even $^4$He is insoluble in the limit $T\rightarrow 0$, with a solubility of order $X_4\sim 10^{-6}$ at $T=2\,\mbox{mK}$~\cite{pet92}.

\begin{figure}[htbp] 
   \centering
   \includegraphics[width=0.7\textwidth]{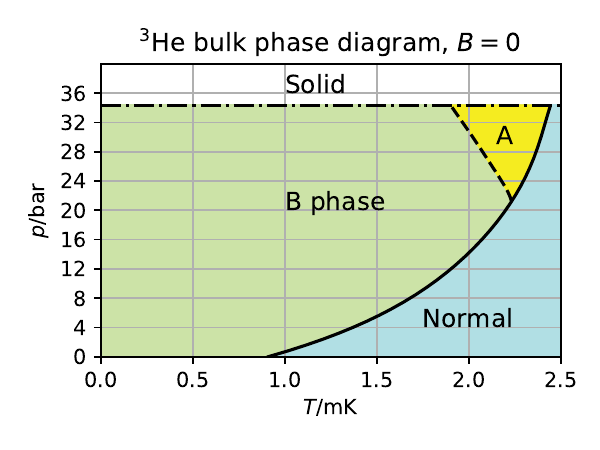} 
   \caption{Phase diagram of bulk \hethree\ at low temperatures and zero magnetic field.}
   \label{f:PhaDia}
\end{figure}

At zero magnetic field, A phase is stable in a small wedge in the plane of temperature $T$ and pressure $P$ below the superfluid critical temperature (see Fig. \ref{f:PhaDia}), but above the polycritical pressure $\PPCP \simeq 21$ bar. As the system is cooled further, A phase becomes metastable below a temperature $\TAB$.  Homogeneous nucleation theory predicts that the lifetime of the A phase at any temperature or pressure is enormous, far longer than the age of the universe.  Yet the transition is observed to happen within in a few hours, even in cells with smooth walls \cite{schiffer1992strong,schiffer1995nucleation}. Thus homogeneous nucleation theory fails dramatically, raising questions as to its extension to cosmological phase transitions.

The QUEST-DMC (Quantum-Enhanced Superfluid Technologies for Dark Matter and Cosmology) collaboration was set up under UK Quantum Technology for Fundamental Physics programme to tackle the nucleation problem using new techniques, both experimental and theoretical, which have become available since the experiments performed in the 1990s.  In this article we report on progress in building simulation code to investigate the dynamics of the order parameter of the system, using established field theory of superfluid $^3$He, and tools and techniques borrowed from cosmological simulations of phase transitions (see e.g. Ref.~\cite{SaulsMizushimaPhysRevB2017} on time-dependent Ginzburg-Landau (TDGL) theory, Ref.~\cite{Hindmarsh:2017gnf} on cosmological phase transitions, and Ref.~\cite{Lizarraga:2016onn,Hindmarsh:2019csc} on topological defects.

\section{Cosmological phase transitions and GWs}

At early times in its evolution the Universe was very close to thermal equilibrium, as the perfect blackbody spectrum of the Cosmic Microwave Background radiation demonstrates \cite{Fixsen:1996nj}. The earlier the time, the higher the temperature, and at very high temperatures the state of matter in the early Universe must change.  At temperatures (thermal energies) higher than around 100 MeV, reached at about 0.1 ms after the Big Bang, the theory of the strong interactions - Quantum Chromodynamics (QCD) - predicts that quarks and gluons inside nucleons are liberated.  At still higher temperatures, around 100 GeV, the 
average value of the Higgs field is predicted to vanish. Elementary particles are then massless, and the electroweak symmetry between the photons, W and Z bosons restored. 

If an early Universe phase transition were of first order the consequences would be very interesting.  The development of homogeneous nucleation theory in quantum field theory \cite{Coleman:1977py, Linde:1980tt},  combined with relativistic combustion theory \cite{landau2013fluid, Steinhardt:1981ct} builds a picture of a transition proceeding by nucleation of bubbles of the low temperature phase by thermal or quantum fluctuations, followed by rapid expansion, which transfers some of the latent heat of the system into motion of the plasma.  The resulting shear stresses generate gravitational waves \cite{Witten:1984rs, Hogan:1986qda}. 
One can estimate that these gravitational waves would be of frequencies observable by pulsar timing arrays (QCD transition, nHz) and space-based interferometers with million-km baselines (electroweak transition, mHz). Another important consequence of a first order electroweak phase transition is that it supplies one of the prerequisites for the dynamical generation of the baryon (matter-antimatter) asymmetry in the Universe: namely departure from thermal equilibrium \cite{sakharov1967violation, Kuzmin:1985mm}.

The first calculations of the free energy of the Standard Model (SM) \cite{Kirzhnits:1972iw, Kirzhnits:1972ut}  indicated that gauge field theories like the electroweak theory have a first order transition, albeit with rather small latent heat \cite{Enqvist:1991xw}.  Early calculations with lattice quantum field theory also showed that gluons confine in a first order phase transition \cite{Fukugita:1989yw,Boyd:1996bx}.  

Further investigation of the Standard Model transitions has shown that they are both cross-overs \cite{Borsanyi:2016ksw, Kajantie:1996mn, Kajantie:1996qd}.  All thermodynamic quantities evolve smoothly with temperature, and while there are peaks in various susceptibilities, there is no possibility that the Universe became stuck in a metastable state.  
The Universe of the Standard Model stays very close to thermal equilibrium, as the particle scattering rate is many orders of magnitude higher than the rate of change of temperature due to expansion.

This is perhaps disappointing.  However, the majority of particle physicists are convinced that the Standard Model is not the ultimate description of matter and interactions.  Apart from lacking a description of gravity, the SM has no explanation for dark matter or the baryon asymmetry (see e.g. \cite{Cline:2018fuq,Croon:2023zay} for pedagogical reviews).  Moreover, calculations of the vacuum fluctuations of the Standard Model particles indicate that the current magnitude of the Higgs field, 174 GeV, is not the lowest energy state \cite{lindner1986implications,Degrassi:2012ry}.  A related puzzle is what determines the magnitude of the Higgs field, and why it is so different from the fundamental mass scale set by gravity, $10^{19}$ GeV.

The problems with the Standard Model motivate extending it. The study of extensions to the Standard Model is known as BSM (beyond the Standard Model) physics and it emerges that 
a first order phase transition in the early Universe is very much a possibility in Standard Model extensions \cite{Caprini:2019egz}. The search for gravitational waves from the early Universe then becomes a search for BSM physics. There follow some major questions:  how to observe these gravitational waves, and how to calculate their spectrum.  Both have become very active areas in the intersection between particle physics and cosmology, and now ultra-low temperature physics.

The simulations and modelling of first order transitions in the early Universe 
(see \cite{Hindmarsh:2020hop} for a review) have shown that the gravitational wave spectrum depends mainly on a handful of thermodynamic quantities: the critical temperature, the transition rate, the latent heat, the phase boundary terminal velocity, and the sound speeds in the two phases.  The most difficult to calculate accurately are the transition rate $\beta$ (the inverse lifetime of the metastable phase) and the phase boundary speed $\vw$, as they are both non-equilibrium quantities.  The amplitude and shape of the gravitational wave power spectrum are quite sensitive to these parameters.  For example, the peak frequency is proportional to the ratio $\vw/\be$ and the transition temperature. 

Calculations of the transition rate $\be$ are based on the homogeneous nucleation theory of Langer \cite{langer1969statistical}, a formalisation of the Cahn-Hilliard theory \cite{cahn1959free}. It was introduced into quantum field theory at zero temperature by Coleman \cite{Coleman:1977py}, and at non-zero temperature by Linde \cite{Linde:1980tt} (see \cite{Berera:2019uyp} for a discussion of the theory). There has been recent progress with perturbative calculations of the rate parameter $\be$ \cite{Gould:2021ccf}, and it 
 has also been calculated non-perturbatively using numerical lattice simulations of Standard Model-like gauge-Higgs system \cite{Moore:2000jw,Gould:2022ran}. Calculations of the wall speed are based on modelling of the plasma in terms of quasi-stable particles and the Boltzmann equation \cite{Liu:1992tn,Moore:1995si}. Here, too, progress can be made with numerical methods \cite{DeCurtis:2022hlx,Laurent:2022jrs}. Both of these frameworks have direct analogies in superfluid \hethree, which offers the opportunity to test and further develop the theory behind the gravitational wave calculations.  

\section{The A-B nucleation puzzle}
 
According to the homogeneous nucleation theory of Cahn and Hilliard \cite{cahn1959free} and Langer \cite{langer1969statistical}, a metastable system makes the transition to the stable state with lower free energy via in small region, which is nevertheless large enough for the pressure difference between the interior and exterior to overcome the surface tension of the boundary between. In the bulk, such a region with lowest energy is spherical: the critical droplet or bubble. The critical bubble provides the route through the space of order parameter configurations to the ground state.

The thermal activation rate per unit volume is 
\ben\label{eq-nucleation_rate}
\Gamma(T,P) = \la_\text{a} n_\text{a} e^{- \Ec/\kB T},
\een
where $\la_\text{a}$ is the attempt frequency, the rate at which the system tries to get over the barrier between the metastable state and ground state, $n_\text{a}$ is the number density of regions in which an attempt can be made, $\Ec$ is the energy barrier, and $T$ is the bath temperature. The attempt frequency and density are set by microscopic dynamics of the order parameter and are difficult to calculate.  This calculation has recently been automated for a single-component order parameter \cite{Ekstedt:2023sqc}, but the rate evaluation for the 18-component order parameter of \hethree\ is far more of a challenging. However, the pair correlation length, $\xi$, provides an estimate for the maximum density, $n_\text{a} \sim \xi^{-3}$. The attempt rate is governed by a combination of inertial dynamics and diffusion over the barrier \cite{langer1969statistical,hanggi1990reaction,Berera:2019uyp}.

In the case of superfluid \hethree, the two phases in question are the A and B phases separated by the first-order transition line $T_{AB}(p)$ (see Fig.~\ref{f:PhaDia}). In zero magnetic field, above the polycritical point pressure $\PPCP\simeq 21$ bar, the A phase is stable below the superfluid critical temperature $\Tc$ and $\TAB$, where the B phase takes over as the phase with lower free energy. 
Both phases belong to the spin-triplet ($S=1$), p-wave ($L=1$) manifold of pairing states defined by the macroscopic  amplitude of fermion pairs, $\langle\psi_{\bp,a}\psi_{-\bp,b}\rangle$, where $\bp$ is the relative momentum of the pair of orbiting $^3$He fermions while $a$ and $b$ are the spin projections ($\uparrow$ or $\downarrow$) of the fermions comprising the pair. The corresponding mean-field pairing self energy, $\Delta_{ab}(\bp) = g\,\langle\psi_{\bp,a}\psi_{-\bp,b}\rangle$, where $g$ is the attractive pairing interaction in the spin-triplet, p-wave Cooper channel. These amplitudes are the elements of a symmetric $2\times 2$ matrix order parameter, $\widehat\Delta(\bp)=i\vec{\sigma}\sigma_y\cdot\vec{d}(\bp)$, where $i\vec\sigma\sigma_y$ are the symmetric Pauli matrices, and the three-component spin vector, $\vec{d}(\bp)$, is in general a linear superposition of the p-wave basis functions for momenta restricted to the Fermi surface.
Thus, 
\begin{equation}
\widehat\Delta(\bp) =(i\sigma_{\alpha}\sigma_y)\,A_{\alpha i}\,(\hat\bp)_i
\,,
\end{equation} 
is parametrized by a $3\times 3$ complex matrix, $A_{\alpha i}$, that transforms as a vector with respect to index $\alpha$ under spin rotations, and, separately, as a vector with respect to index $i$ under orbital rotations.\footnote{We follow the notation of Ref.~\cite{mer73} for the form of the spin-triplet, p-wave order parameter and Ginzburg-Landau (GL) functional. See also Vollhardt and W\"olfle~\cite{vollhardt90} for a pedagogical development of the same.} This representation for the order parameter provides us with an $S=1$, $L=1$ basis for an irreducible representation of the maximal symmetry group of normal $^3$He,
\begin{equation} \label{eq:bulk_symmetry}
G = SO(3)_{\mathbf{L}} \times SO(3)_{\mathbf{S}} \times U(1)_{\mathbf{N}} \times T \times C  \times P,
\end{equation}
which includes $SO(3)_{\mathbf{L}}$ rotations in three-dimensional space, $SO(3)_{\mathbf{S}}$ rotations in the spin space, and the global phase transformation group $U(1)_{\mathbf{N}}$, as well as discrete symmetries $T$, $C$ and $P$, where $T$ is time-reversal symmetry, $C$ is the particle-hole symmetry, and $P$ is parity symmetry. The subscripts $\mathbf{L}$, $\mathbf{S}$ refer to the generators for the rotation groups while $\mathbf{N}$ is the number operator which is the generator for changes in phase.

The bulk B-phase, which minimizes the bulk free energy over most of the pressure-temperature plane below $T_c$, is the ``isotropic'' state defined by 
\begin{equation}
    A_{\alpha i}^{\text{B}} = \frac{1}{\sqrt{3}}\Delta_{\text{B}}\,\delta_{\alpha i}
    \,,
\end{equation}  
with residual symmetry $H_{\text{B}} = SO(3)_{\mathbf{L}+\mathbf{S}} \times T$, i.e. the B-phase is time-reversal invariant and invariant under joint rotations of spin and orbital components of the order parameter. This state was shown by Balian and Werthamer to be the ground state for spin-triplet, p-wave pairing in the weak-coupling limit~\cite{bal63}. 
The high degree of symmetry of the B-phase implies a large continuous degeneracy space,
\ben
R_{\text{B}}=U(1)_{\text{N}}\times SO(3)_{\mathbf{L}-\mathbf{S}}
\,,
\een
corresponding to the choice of phase of the order parameter defined by the elements of $U(1)_{\text{N}}$, as well as the relative orientation of the spin and orbital coordinates of the spin-triplet, p-wave Cooper pairs, represented by $SO(3)_{\mathbf{L}-\mathbf{S}}$. 
As a result the class of degenerate B-phase order parameters is,
\begin{equation}
A_{\alpha i}^{\text{B}}
=\frac{1}{\sqrt{3}}\Delta_{\text{B}}\,e^{i\varphi}\,R_{\alpha i}[\vartheta\hat\bn]
\,,
\end{equation}  
where $\varphi$ is the global phase and $R_{\alpha i}[\vartheta\hat\bn]$ is an orthogonal matrix defining a rotation of the spin and orbital coordinates by angle $\vartheta$ about the direction $\hat\bn$. Thus, 
there are 4 continuous degeneracy parameters, and hence 4 gapless Nambu-Goldstone (NG) modes. The phase mode is realized as collisionless sound in superfluid $^3$He-B and plays a central role in observations of the spectrum of Higgs modes~\cite{hal90,sau00a}. There are three spin-orbit modes that are key signatures of spin-triplet pairing in NMR spectroscopy of $^3$He~\cite{LeggettRMP1975}. Nuclear dipolar interactions and Zeeman energies in an external magnetic field partially lift the degeneracy of these modes, opening small gaps providing a novel example of the Light Higgs scenario~\cite{zav16}.

The stability of the A-phase at high pressure and temperatures relatively near to $T_c$ results from corrections to weak-coupling BCS theory that become sufficiently large at high pressures to stabilize an equal-spin pairing (ESP) state that also spontaneously breaks time-reversal symmetry with an order parameter of the form,
\begin{equation}
    A_{\alpha i}^{\text{A}} = \Delta_{\text{A}}\,\hat{\bd}_{\alpha}\,\left(\hat{\bm}_i + i\,\hat{\bn}_i\right)/\sqrt{2}
    \,,
\end{equation}  
where $\hat\bd$ is a real unit vector in spin space along which the A-phase Cooper pairs have zero spin projection. The orthonormal unit vectors $\hat\bm$ and $\hat\bn$ combined with the relative phase of $\pi/2$ define orbital motion of the A-phase Cooper pairs with orbital angular momentum $+\hbar$ per Cooper pair along the axis $\hat\bl = \hat\bm\times\hat\bn$. This axis is chiral and it highlights both broken mirror symmetry and broken time-reversal symmetry by the A-phase. The latter allows for a macroscopic ground state angular momentum predicted to be, $\mathcal{L}_z = {N}\hbar/{2}$ for a system with $N$ $^3$He atoms.~\footnote{See Ref.~\cite{sau11} for a discussion of the connection between the topological edge states, the edge currents and $\mathcal{L}_z$ as well as review of theoretical literature on the ground state angular momentum.}
The corresponding residual symmetry group is then,
$H_{\text{A}} = SO(2)_{\bd} \times U(1)_{\mathbf{L}_z-\mathbf{N}} \times Z_2$, where 
$SO(2)_{\bd}$ is the group of rotations in spin space about the axis $\hat\bd$. The A phase breaks orbital rotation symmetry as well as global gauge symmetry, however a rotation by any angle about the chiral axis can be undone with an appropriately chose element of $U(1)_{\mathbf{N}}$ which leads to the residual gauge-rotation symmetry of the A-phase defined by $U(1)_{\mathbf{L}_z-\mathbf{N}}$. The residual discrete $Z_2$ symmetry results from the combination of time-reversal and mirror reflection in a plane containing the chiral axis. This symmetry allows for the remarkable transport properties of the A phase including the anomalous Hall effect that was reported for electrons moving in $^3$He-A driven by an electric field perpendicular to the chiral axis~\cite{ike13,she16}.

The A phase is also endowed with a large degeneracy space, in this case by the combined degeneracy in the orientation of the spin direction $\hat\bd$ on the surface of a unit sphere, $S^2$, and the orientation of the orbital triad, $\{\hat\bm,\hat\bn,\hat\bl\}$, which is the group of rotations in $3$-space, $SO(3)$. However, the combined transformations: $\hat\bd\rightarrow -\hat\bd$ and $\left(\hat{\bm} + i\,\hat{\bn}\right)\rightarrow - \left(\hat{\bm} + i\,\hat{\bn}\right)$ is a discrete symmetry of the A-phase ($Z_2^{'}$), and thus the degeneracy space excludes these combined changes of sign such that~\cite{volovik1992exotic}
\ben
R_{\text{A}}=S^2 \times SO(3)/Z_2^{'}
\,.
\een
The continuous degeneracy space implies the existence of 5 Nambu-Goldstone modes, 2 spin wave modes, 2 orbital wave modes and the sound mode. It is also worth noting that the $Z_2^{'}$ symmetry is directly related to topologically stable half quantum vortices originally predicted by Volovik and Mineev for ESP states~\cite{vol76,salomaa1987quantized}, and which were recently discovered in NMR spectroscopy of the ESP polar phase of $^3$He under rotation~\cite{autti2016observation}.

In simulating nonequilibrium dynamics and nucleation processes the low energy excitations - Goldstone and pseudo-Goldstone modes - are expected to play an important role in transporting mass, energy and magnetization.
Another notable fact is that the residual symmetry group of the B phase is not a sub-group of $H_{\text{A}}$, and thus the phase transition is necessarily first order. Thus, to nucleate the B phase from the homogeneous A-phase requires deviations or fluctuations of the order parameter from the local equilibrium A phase that incur an energy barrier, inhibiting nucleation and allowing for supercooling of the A phase below $T_{\text{AB}}$.
Cooling below $\TAB$ at pressures above $\PPCP$ puts the superfluid into a metastable state with free energy excess $\De \fAB = \fA - \fB$, where $\fA$ and $\fB$ are the condensation energy densities of the A- and B-phases, which can be determined by integrating the measured specific heats from $T_c$ to the relevant temperature below $T_c$. 

The path in order parameter space that minimizes the energy cost of a domain wall (DW) separating the A and B phases, $A_{\alpha i}^{\text{A}}\rightarrow A_{\alpha i}^{\text{DW}}(x)\rightarrow A_{\alpha i}^{\text{B}}$, for a bubble of radius $R$ of $^3$He-B embedded in $^3$He-A plays a key role in the theory of nucleation of the  -phase in supercooled A phase. The surface energy of the A-B interface was measured at high pressure by Osheroff and Cross~\cite{osheroff1977interfacial}, and at low pressure and high magnetic field by Bartkowiak et al \cite{bartkowiak2004interfacial}. Theoretical calculations using  GL theory \cite{kaul1980surface,schopohl1987spatial,thuneberg1991b} give $\siAB \propto \xi \fB$, where $\xi$ is the Ginzburg-Landau coherence length, with proportionality constant close to 1 in good agreement with experiment.

To estimate the energy and radius of the critical bubble in the thin-wall approximation ($R\gg\xi$) we express the total energy of a bubble of B-phase embedded in metastable A-phase as a sum of the gain in condensation energy proportional to $-\De\fAB$, and the cost in surface energy, proportional to $\siAB$. Both depend on temperature and pressure. Thus, 
\ben
E(R) = 4\pi R^2 \siAB - \frac{4\pi}{3}R^3 \De \fAB
\,.
\een
The critical bubble is determined by the condition $E'(R) = 0$, representing a spherical bubble poised between expansion and contraction. The radius and energy of the critical bubble are then
\ben
\Rc = \frac{2\siAB}{\De \fAB}, \quad \Ec = \frac{16\pi}{3} \frac{\siAB^3}{\De \fAB^2},
\een
and the critical bubble free energy is 
\ben
\Ec \simeq \ep |\fB|\xi^3 \left(\De\fAB/|\fB|\right)^{-2}, 
\een
where $\ep \simeq 10$. The Ginzburg-Landau coherence length near the critical temperature is $\xi = \xiGL(1 - T/\Tc)^{-1/2}$, where $\xiGL=\sqrt{7\ze(3)/20}\,\xi_0$ and $\xi_0 = \hbar\vF/2\pi\kB\Tc$ is the Cooper pair correlation length in the ballistic limit~\cite{wim16}. 
The order of magnitude of the condensation energy density is set by the density of states at the Fermi surface $N(0) = m^{*}k_f/(2 \pi^{2} \hbar^{2})$ and the critical temperature according to 
$\fB \sim - N(0) (\kB \Tc )^2(1 - T/\Tc)^{2}$. Hence 
\ben
\label{e:CriBubEne}
\Ec/\kB T \sim \Gi \, ({\Tc}/{T}) \left( 1 - T/\Tc \right)^{1/2}  \left(\De\fAB/|\fB|\right)^{-2}.
\een
where $\Gi = N(0) \xi_0^3 (\kB \Tc )$ is the Ginzburg number, which takes values between 800 at zero pressure and 2900 at melting pressure, about $34$ bar.  A plausible estimate for the order of magnitude of the length scale in the attempt density is the Cooper pair correlation length $\xi_0$, which takes values in the range 16 nm to 77  nm.  An upper bound the attempt rate is $\vF/\xi_0$, where the Fermi velocity $\vF$ is in the range $30 \, \text{m}\,\text{s}^{-1}$ to $60 \, \text{m}\,\text{s}^{-1}$.

The last factor in Eq.~\eqref{e:CriBubEne} diverges as $(1 - T/\TAB)^{-2}$ as the A-B equilibrium line is approached, but even without the divergent factor the size of the Ginzburg number already ensures that the exponential $e^{-E_c/\kb T}$ in the nucleation rate, Eq.~\eqref{eq-nucleation_rate}, overwhelms the attempt frequency in an experimentally accessible volume, leading to an estimate of the lifetime of the metastable A-phase that vastly exceeds the current age of the universe. Thus, as pointed out early after the discovery of the superfluid phases~\cite{kaul1980surface,leggett1984nucleation} classical nucleation theory predicts that the superfluid $^3$He should remain in the metastable A-phase indefinitely in any experiment that can be realistically conceived.
However, experimental investigations of supercooled $^3$He-A all show that the B phase nucleates on timescales of seconds to hours suggesting another mechanism is responsible for the nucleation of $^3$He-B~\cite{kleinberg1974supercooling,hakonen1985comment,schiffer1992strong,schiffer1995nucleation,schiffer1995nucleation,schiffer1995PLTPnucleation,zhelev2017ab,lotnyk2021path,tian2023supercooling}.

\section{\label{scenerios} Explanations for the nucleation puzzle}

Proposed explanations for the puzzle are many and varied~\cite{leggett1984nucleation,hong1988q,leggett1990nucleation,schiffer1995PLTPnucleation,VolovikCzechJP1996, bunkov1998cosmological,balibar2000comments,tye2011resonant,tian2023supercooling}. The leading contenders consider that nucleation in bulk, metastable superfluid $^3$He-A is caused by energy injection by cosmic-ray muons or another energetic  particle. In the ``Baked Alaska'' scenario proposed by Leggett~\cite{leggett1984nucleation} the energy deposition breaks Cooper pairs creating a local region of ``hot'' quasiparticles surrounded by the cold metastable A phase. A shell of energetic quasiparticles expands, driving the system locally normal,  behind which the superfluid returns to a temperature below $T_c$, allowing the B-phase to nucleate with measurable probability. At this point the size of the shell must be larger than the critical bubble size, otherwise the surface tension of the A-B phase boundary will overwhelm the pressure difference and the B-phase bubble will collapse. 
Support for such a scenario of local heating nucleating the B-phase is reported by Schiffer et al. where it was shown that 764 keV neutrons as well as MeV $\gamma$ rays from $^{60}$Co stimulate the A-B transition~\cite{schiffer1992strong}.

A particularly interesting consequence of a second-order symmetry-breaking phase transition such as the normal-superfluid transition in \hethree, was pointed out by Kibble in the context of phase transitions in the early Universe: the generation of topological defects~\cite{Kibble:1976sj}. He showed how to predict the type of defect on the basis of the topology of the manifold of equilibrium states and the defect density on the basis of estimates of the correlation length as the universe cools through a continuous phase transition. The density estimate was later updated by Zurek who made the explicit link to defect formation in rapid quenches in superfluid $^4$He~\cite{Zurek:1985qw,Zurek:1996sj}.  Experiments to investigate vortex formation in superfluid $^4$He \cite{hendry1994generation,dodd1998nonappearance} proved inconclusive, but experiments in the B phase of superfluid $^3$He \cite{bauerle1996laboratory,ruutu1996vortex} demonstrated spontaneous vortex generation in rapid quenches consistent with the Kibble-Zurek mechanism.

The ``Cosmological Scenario'' for A-B nucleation proposed by 
Volovik and Kibble ~\cite{VolovikCzechJP1996,kibble1997phase} envisages that energy is transported rapidly out of the injection region by thermal diffusion, and that the front where the quasiparticle temperature goes below $\Tc$ is swiftly followed by another front where it goes below $\TAB$. The rapidity of the quench inside the energy deposition region suggests that it contains causally disconnected regions of local order, which evolve either into the A or B phases
according to the Kibble-Zurek scenario.  
In this case a complex region of multiple phases, separated by domain walls, emerges. If a large enough B phase region has formed, it will expand and eventually take over the condensate. 
Moreover, different types of topological defects such as domain walls between B-phases \cite{SalomaaPRB1988}, as well as vortices can be generated in the energy deposition region. Indeed vortices have been detected in the Helsinki group's experiments injecting energy with neutrons \cite{ruutu1996vortex}.  
There has been discussion between the authors of the competing models \cite{schiffer1999physical,bunkov1999bunkov}, but as yet no consensus.

It has also been pointed out that nucleation may be seeded by complex order parameter configurations at the boundaries of an experimental cell, either by surface roughness, or by topological singularities \cite{mer77,leggett1990nucleation}.  Nucleation at rough boundaries has been argued to fit the data of Hakonen et al.~\cite{hakonen1985comment,balibar2000comments}, where nucleation occurred close to a ``catastrophe'' line in the $(T,P)$ plane with a characteristically peaked temperature distribution.  The catastrophe line would be different for each experiment, as it would depend on the details of the boundaries, particularly in the heat exchanger, where complex surfaces of many square metres in area are found. Recently, it has been shown that the position of the catastrophe line also depends on the path in the $(T,P)$ plane taken when cooling through the A phase into the metastability region \cite{tian2023supercooling}.  This was explained in terms of a model of complex order parameter configurations acting as B-phase seeds in small cavities, principally the heat exchanger.  

More exotic explanations have been put forward.  Non-topological order parameter configurations known as Q-balls \cite{coleman1985q} have been proposed as an alternative to the critical bubble as the route from the A phase to the B phase \cite{hong1988q}. Such objects have been experimentally detected and studied \cite{autti2018propagation}.   It has also been proposed that resonant tunnelling, a quantum-mechanical phenomenon where quantum tunnelling can proceed via a classically allowed intermediate state of the same energy, could be at play in superfluid \hethree\ as well \cite{tye2011resonant}.  However, such classically allowed states were shown not to exist in the quantum field theory of a single scalar field \cite{copeland2008no}, so the existence of resonant tunnelling for the multicomponent order parameter of \hethree\ is not clear.

\section{New experiments on A-B nucleation}

In order to study nucleation in the bulk superfluid, one would like to eliminate or control the effect of the container walls. The first imperative is to isolate the metastable superfluid from the heat exchanger, which contains a large rough area in contact with the superfluid. This can be done in two ways. 
One method is to utilize the Zeeman energy of the ESP A phase in which case its equilibrium bulk free energy is reduced in a magnetic field.  Above about 0.6 T, the A phase is the stable superfluid phase over the whole $(T,P)$ plane \cite{hahn1994phase,hahn1993phdthesis}. By placing two opposing magnets close to each other, it is possible to create a region of low magnetic field, where the superfluid is in the metastable A phase, surrounded by higher field, where the A phase is stable \cite{bradley2006levitated}. In this configuration the phase transition happens in the low-field region, well away from the walls of the container. 

A new set of experiments use engineered cells and surfaces to confine $^3$He into multiple chambers of different heights within a single experimental cell to study A-B nucleation. 
Container walls generally lead to pair breaking and distortion of the order parameter near the wall. The magnitude of pair-breaking depends on the atomic scale properties of the wall~\cite{ambegaokar1974landau,freeman1988size,vorontsov2003thermodynamic}. In general pairing of states with orbital angular momentum in the plane of a surface are suppressed.  
If the wall is smooth, e.g. by pre-plating with superfluid $^4$He, quasiparticles reflect specularly, and the in-plane orbital states are unaffected. 
This is the case for the A phase with the chiral axis aligned normal to the wall; pair-breaking is suppressed and thus the the A-phase order parameter survives all the way to the wall. The B phase, on the other hand, is modified by pair-breaking of the orbital component normal to the wall. Its order parameter is distorted towards the planar phase at the wall. The planar distorted B phase extends into the bulk over a distance of a few coherence lengths. In the case of an atomically rough surface pair-breaking occurs for all orbital components, in which case both the A and B phase are suppressed near the wall by diffuse scattering. Thus, the ideal geometry is a slab that stabilizes the A phase with minimal pair-breaking, i.e. by $^4$He pre-plating, below $T_c$, but is thick enough to support the B-phase at lower temperatures as shown for example in Fig.~3 of Ref.~\cite{vorontsov2003thermodynamic}.
Indeed, experiments in thin nanofabricated cavities show that A phase can be stabilised at any pressure over a range of temperatures \cite{levitin2013phase,HeikkinenNatureComm2019}, and is the stable superfluid phase that onsets at the critical temperature.  

These facts motivated the construction of the experimental cell used in the QUEST-DMC experiments which consists of 5 superfluid ``lakes'' of $^3$He, each being approximately 7 $\mu$m in depth, surrounded by shallower regions of $\sim$70 nm, in which the $^3$He is forced to be either in the normal phase (diffuse scattering) or {\it only} the A phase (specular scattering). This design allows for the study of the A-B transition in multiple regions of metastable A phase in lakes of various volumes. The results of the intital studies in this geometry are reported elsewhere \cite{Heikkinen:2024any}.
The ultimate goal is study intrinsic A-B nucleation by understanding, controlling and potentially eliminating extrinsic nucleation related to boundaries, defects, and particles depositing energy in $^3$He. 

\section{\label{KZVTDGL} Simulations of nonequilibrium phase transitions}

In order to develop a deeper understanding of nucleation mechanisms for phase transitions, the QUEST-DMC theory effort is developing computational tools to simulate nonequilibrium dynamics for the A-B transition of metastable $^3$He-A. This is a technically challenging problem involving the dynamics of a bosonic field defined on a multi-dimensional order parameter space coupled to excitations of the underlying fermionic vacuum. In many cases the dynamics occurs under conditions that are far from equilibrium, particularly for nucleation generated by localized energy deposition. Here we discuss simulations based on dynamics described by TDGL theory based on the extension of the strong-coupling GL functional by Wiman and Sauls~\cite{wim15,wim16} that captures the A-B transition and extends the GL theory to temperatures below $T_{\text{AB}}$~\cite{reg20}.

\subsection{Time-dependent Ginzburg-Landau theory}

Time-dependent Ginzburg-Landau (TDGL) equations have long been studied in the context of the nonequilibrium dynamics of superconductors, particularly for superconductors in the ``dirty'' limit, $\hbar/\tau\gg\Delta$, where $\tau$ is the mean scattering time for unbound fermionic quasiparticles (see e.g. Kopnin's review~\cite{kop02}). These equations have been also been studied in the context of Kibble-Zurek quench dynamics and normal-superfluid boundary propagation for $U(1)$ superfluids by several authors~\cite{AransonKupninVinokurPRL1999,AransonKupninVinokurPRB2001,KopninThunebergPRL1999}. TDGL equations for superconductors and superfluids in the clean limit were developed early on by Abrahams and Tsuneto for a $U(1)$ superconductor starting from an expansion of the nonequilibrium mean-field equations for the order parameter~\cite{abrahams1966time} (see also Ref.~\cite{SadeMeloPRL1993}), and for superfluid $^3$He by Kleinert~\cite{kleinert1978collective}. Below we formulate the dynamics as a bosonic field theory for superfluid $^3$He with dissipation from the excitations of the underlying fermionic vacuum.

The space-time evolution of the bosonic field describing Cooper pairs in bulk $^3$He, $A_{\alpha i}(\br,t)$, is governed by field equations obtained from the TDGL Lagrangian, $L=K-U$, where 
\ben\label{eq-TDGL_kinetic_term}
K = \int dV\, \tau_0\, \Tr{\dot{A} {\dot {A}}^\dagger}
\,,
\een
is the kinetic energy associated with temporal fluctuations of the field $A(\br,t)$ with $\dot{A}=\partial{A}/\partial t$ and $\tau_0$ the inertia of the field which determines the dispersion of the bosonic modes of the superfluid phases~\cite{SaulsMizushimaPhysRevB2017}.
The potential energy functional is defined by the GL free energy functional, which includes a second-order invariant,
\begin{equation}
U_2[A] = \int dV\,\alpha\,\Tr{A\,A^{\dag}}
\,,
\end{equation}
that controls the phase transition to the broken symmetry phase. For $\alpha >0$ the equilibrium state is the symmetric normal Fermi-liquid, while for $\alpha<0$ the equilibrium state spontaneously breaks the symmetry, with a finite bosonic amplitude $A$.
The broken symmetry equilibrium state is determined by the fourth-order interactions of the bosonic field,
\begin{equation}
U_4[A] = \int dV\,\sum_{p=1}^{5}\beta_{p\,}\,u_{p}(A)
\,,
\end{equation}
where the five linearly independent fourth-order invariants of the maximal symmetry group $G$ for $^3$He in Eq.~\eqref{eq:bulk_symmetry} are
\begin{eqnarray}
u_1 &=& |\Tr{A A^T}|^2 
\,,\quad
u_2 = \Tr{A A^\dagger}^2 
\nonumber\\
u_3 &=& \Tr{A A^T A^* A^\dagger} 
\,,\quad
u_4 = \Tr{A A^\dagger A A^\dagger} 
\,,\quad
u_5 = \Tr{A A^\dagger A^* A^T }
\,.
\end{eqnarray}
Spatial gradients of the bosonic field also play a central role in the dynamics and contribute to the effective potential in the form of supercurrents, textural bending energies and deformations of the order parameter near the cores of singular topological defects and boundaries,
\begin{equation}\label{eq-gradient_energy}
U_{\partial}[A] = \int dV\,\sum_{m=1}^{3}\,K_{m\,}\,v_{m}(\partial A)
\,,
\end{equation}
where the three linearly independent leading order gradient energies are
\begin{eqnarray}
v_1 &=& \pa_k A_{\al j}\pa_k A^*_{\al j}
\,,\quad
v_2 = \pa_j A_{\al j}\pa_k A^*_{\al k} 
\,,\quad
v_3 = \pa_k A_{\al j}\pa_j A^*_{\al k}
\,.
\end{eqnarray}
Thus, the Lagrangian density for the bosonic fields takes the form,
\begin{eqnarray}\label{eq-Lagrangian}
\mcL &=&
\tau_0\,\Tr{\dot A\dot A^{\dag}}
-
\alpha\,\Tr{A\,A^{\dag}}
-
\sum_{p=1}^{5}\beta_{p\,}u_{p}(A)
-
\sum_{m=1}^{3}\,K_{m}\,v_{m}(\partial A)
\,,
\end{eqnarray}
where $\tau_0$ is the effective inertia for Cooper pair fluctuations.
This Lagrangian respects the maximal symmetry group $G$ of $^3$He in Eq.~\eqref{eq:bulk_symmetry}. 
Weak violation of particle-hole symmetry by the parent Fermi-liquid allows for an additional invariant in the Lagrangian that is first-order in $\partial_t A$,
\ben
K_{\Gamma} = i\Gamma \left[\Tr{\dot{A} A^\dagger} - \Tr{A \dot{A}^\dagger}\right]
\label{CCiolationTerm}
\,.
\een 
However, this C-violating term is expected to be small; thus we neglect it in the dynamical simulations that follow. In contrast we retain the dissipative term that is first order in $\partial_t A$ that arises from coupling of the thermal bath of fermionic excitations to nonequilibrium states of the bosonic field as discussed in Sec.~\ref{sec-TDGL_Dissipation}.

The Lagrangian (\ref{eq-Lagrangian}) generates the dynamical equations for the bosonic excitations of superfluid $^3$He~\cite{SaulsMizushimaPhysRevB2017}.
The bosonic field theory is significant as it provides an understanding of fundamental dynamical features resulting from spontaneous symmetry breaking in condensed matter and quantum field theories. A good example is Nambu's fermion-boson mass relations~\cite{nam85} for the broad class of Nambu/Jona-Lasinio field theories~\cite{nam61,nam09}, which includes $^3$He, for mass generation by spontaneous symmetry breaking~\cite{SaulsMizushimaPhysRevB2017,VolovikZubkovJLTP2014}.

The parameters $\al$, $\be_p$ and $K_m$ that define the effective potential are temperature- and pressure-dependent, and can be calculated from the microscopic theory of superfluid $^3$He~\cite{rai76,sau81b,sau81c,wim15,wim19}, have the values  
\ben
\al(T) = \frac{1}{3}N(0)(T/\Tc  - 1),
\een
\ben
\be_p = \be_0\left( b^\text{wc}_p + \frac{T}{\Tc} b^\text{sc}_p\right)\,,\quad p\in\{1,\ldots, 5\}
\label{eq-beta_parameters}
\een
where the $\beta_p$ parameters in the weak-coupling limit are determined by pressure-independent ratios and an overall scale set by
\ben
\be_0 = \frac{7\ze(3)}{80\pi^2} \frac{N(0)}{3 (\kB \Tc)^2} , \quad \{b^\text{wc}_p\} = (-1, 2, 2, 2, -2)
\,.
\een
The strong-coupling corrections to the $\beta_p$ parameters, $b_p^{\text{sc}}$, are calculated based on the leading order corrections to weak-coupling BCS theory as formulated by Rainer and Serene~\cite{rai76}. A key result is that the strong-coupling corrections to the GL functional are determined by the scattering amplitude for normal-state quasiparticles with energies and momenta confined to the Fermi surface. This scattering amplitude also determines the normal-state thermodynamic and transport properties of the normal Fermi liquid phase of $^3$He. This allows us to solve the inverse problem to determine the scattering amplitude from the experimental data for the normal Fermi liquid phase of $^3$He and the heat capacity jumps for the A- and B-phases at $T_c$. This program was carried out and shown to predict the stability of the A-phase above the polycritical pressure as well as the temperature dependence of the gap, thermodynamic potential and heat capacity of the B-phase at low temperatures~\cite{sau81b,sau81c,wim19}. The calculated A-B transition line is in excellent agreement with experimental results as shown in Fig.~8.6 of Ref.~\cite{wim19}.

The other development in strong-coupling theory for $^3$He is the recognition of the importance of the temperature-dependent scaling of the strong-coupling $\beta$ parameters below $T_c$ shown in Eq.~\eqref{eq-beta_parameters}. This scaling is based on the microscopic strong-coupling theory and developed in Refs.~\cite{wim15,wim16}. The temperature and pressure dependence of the strong-coupling corrections captures the A-B transition transition line to good accuracy, and extends the predictive capabilities of the GL theory to temperatures below $T_{\text{AB}}$~\cite{reg20} for pressures above the polycritical pressure. This is essential for developing TDGL theory to study order parameter dynamics in the metastable A phase. In what follows we use the results for  $b^\text{sc}_i(p)$ tabulated in Ref.~\cite{reg20}; these values are slightly different than the more accurate results reported in Ref.~\cite{wim19}, but both sets are comparable in their magnitude and pressure dependences.\footnote{A review of strong-coupling theory, including the effective interactions in liquid $^3$He that give rise to the stability of $^3$He-A and deviations from weak-coupling BCS theory for the thermodynamic properties, will be published in a separate report.}

\begin{figure}[t]
   \begin{center}
   \includegraphics[width=0.7\textwidth]{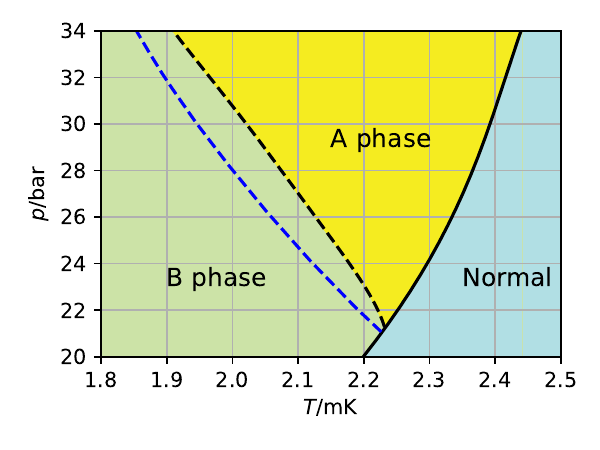} 
   \end{center}
   \caption{Phase diagram of bulk \hethree\ showing the experimental superfluid critical temperature $\Tc$ (black solid line), the A-B equilibrium line  $T_{\text{AB}}$ (black dashed line) \cite{greywall1986he} and the theoretical result for $T_{\text{AB}}$ based on the linear temperature scaling and pressure-dependence of the strong-coupling $\beta$ parameters below $\Tc$ \cite{reg20}.
   \label{fig-Phase_Diagram-TAB}
   }
\end{figure}

The spatial derivative terms in Eq.~\eqref{eq-gradient_energy} determine the energy cost of deformations of the order parameter from its homogenous equilibrium value. All three stiffness coefficients are positive and can be calculated to good approximation in weak-coupling theory,
\begin{equation}
 K_1 = K_2 = K_3 =\frac{7\zeta(3)}{60} N(0) \xi_{0}^{2}
 \,,
\end{equation}
where $\xi_{0}=\hbar v_{\text{F}}/2 \pi k_{B} T_c$ is the zero-temperature Cooper pair correlation length.

The counterpart to the space derivative terms given in Eq.~\eqref{eq-TDGL_kinetic_term} is the kinetic energy which is second-order in time-derivatives. The stiffness to temporal fluctuations of the order parameter in the weak-coupling limit is given by
\begin{equation}
 \tau_0 = \hbar^2\frac{7\zeta(3)}{48} \frac{N(0)}{\pi^2\kb^2 T_c^2}
 \,.
\end{equation}
This result is equal in magnitude to the fourth-order contribution to the linear combination of $\beta$ parameters that determine the gap amplitude for the B-phase, i.e. $\tau_0 = \beta_{\text{B}}=\beta_{12}+\beta_{345}/3$ in weak-coupling theory. 
This result guarantees that the $J=0^+$ Higgs mode has the mass of two fermions at the continuum edge, $M_{0^+}=2\Delta_{\text{B}}$. Based on symmetry grounds it is argued that the relation $\tau_0 = \beta_{\text{B}}$ is preserved to leading order in strong-coupling theory~in~\cite{SaulsMizushimaPhysRevB2017}.

For static, but in general homogeneous superfluid phases, the effective potential,
$U_2 + U_4 + U_{\partial}$,  defines the GL free energy functional, $\mathcal F_{\text{GL}}[A] = \int dV\,f_s[A]$, with 
\ben\label{eq-fGL}
f_s[A] = 
\alpha\,\Tr{A\,A^{\dag}}
+
\sum_{p=1}^{5}\beta_{p\,}u_{p}(A)
+
\sum_{m=1}^{3}\,K_{m}\,v_{m}(\partial A)
\,.
\een
The GL functional is supplemented by boundary conditions for the order parameter $A$ that depend on the geometry and atomic scale properties of the boundary. In the simulations discussed later in this report we compare energy densites relative to that of the homogeneous B-phase,
$f_{\text{B}} = - \alpha^2/4\beta_{\text{B}}$ and we normalize to the energy density scale 
\ben
f_0=\frac{1}{3} N(0) (\kb T_c)^2.
\een

It is worth noting that our formulation of TDGL theory for the space-time dynamics of pure spin-triplet, p-wave superfluid $^3$He can be extended to include attractive, but sub-dominant, spin-triplet, f-wave Bosonic excitations, including the predicted $S=1$, $L=3$, $J=4^{-}$ Higgs mode~\cite{sau81} for which there is experimental evidence from acoustic Faraday rotation of transverse sound~\cite{dav08}. 

The general form of the $S=1$, $L=3$ order parameter is $F_{\alpha;ijk}(\br,t)$ which transforms as a vector under $SO(3)_{S}$ for the index $\alpha$ and as a rank 3 symmetric, traceless tensor under orbital rotations ($SO(3)_{L}$) for the indices $i,j,k$. The leading order contribution to the effective potential is then $\alpha_f(T)\,F_{\alpha;ijk}F^*_{\alpha;ijk}$, where $\alpha_f(T) = \frac{1}{3}\,N(0)\,\left(T/T_{c_f} -1\right)$ where $T_{c_f}$ is the f-wave pairing instability temperature which is a direct measure of the f-wave pairing interaction. For sub-dominant f-wave pairing we have $0 < T_{c_f} < T_c$. There are many new invariants that contribute the extended TDGL functional which can be enumerated using group representation theory. 
Whether or not there is an f-wave condensate depends on the material parameters of the new invariants. 
It may be possible that such a condensate exists in the cores of o-vortices in the B phase, analogous to the p-wave condensate in the core of an Abrikosov (s-wave) vortex \cite{salomaa1987quantized}.
%

Finally we note that symmetry breaking perturbations from the nuclear Zeeman energy in an external magnetic field and the nuclear magnetic dipole-dipole energy can be included in this framework, c.f. Refs.~\cite{miz12a,wim15}, but here we focus on A-B transition in the absence of magnetic field and neglect the weak nuclear dipole-dipole energy. These effects of these perturbations on A-B transition will be discussed in a future work.

\subsection{TDGL equations with dissipation by the fermionic bath}\label{sec-TDGL_Dissipation}

The Euler-Lagrange equations obtained from Eq.~\eqref{eq-Lagrangian} generate the non-dissipative coupled dynamical equations for the $3\times 3$ complex matrix order parameter, $A_{\alpha i}(\br,t)$, for superfluid $^3$He,
\ben \label{TDGLEqns}
\ba{rcl}
\tau_0\ddot A_{\al i} 
&+& 
  \al  A_{\al i} 
 -
 K_1\pa^2 A_{\al i} - (K_2+K_3) \pa_i\pa_j A_{\al j} 
\\
&&+ 2\left[
\be_1 A^*_{\al i} \Tr{A A^T} + 
\be_2 A_{\al i} \Tr{A A^\dagger} \right.
\\
&&\qquad
+ \left.
\be_3 (A A^T A^*)_{\al i} +
\be_4 (A A^\dagger A)_{\al i} + 
\be_5 (A^* A^T A)_{\al i} 
\right] 
= 0
\,.
\ea
\een
These coupled equations involve only the bosonic degrees of freedom with parameters corresponding to the fermionic vacuum in local equilibrium. 
We include thermal fluctuations of the fermionic vacuum via a Langevin source term that couples the thermal fluctuations locally to the bosonic field. Several authors have formulated the dynamics of a bosonic field theory coupled to a thermal bath in terms of a stochastic Langevin source~\cite{ant99,AransonKupninVinokurPRL1999}. 
We add to the right-hand side of Eqs.~\eqref{TDGLEqns} a Gaussian noise source $\noise_{\alpha i}(\br,t)$ with intermediate-time averages $\langle\noise_{\alpha i}(\br,t)\rangle = 0$ and~\footnote{Our formulation is similar that of Ref.~\cite{ant99}. }
\begin{equation}
\langle \noise_{\alpha i}(\br,t)\noise_{\beta j}(\br',t')\rangle = 2\gamma\,\kb T\,\delta_{\alpha\beta}\delta_{ij}\delta(\br-\br')\delta(t-t')
\,.
\end{equation}
The parameter $\gamma$ plays an important role as it leads, via the fluctuation-dissipation theorem, to damping of space-time fluctuations of the bosonic field via an additional dissipative time-derivative term, $\gamma\dot A_{\alpha i}$, on the left-side of Eq.~\eqref{TDGLEqns} that is characteristic of Langevin dynamics. 
Thus, the set of dynamical equations including the dampling and Langevin noise source terms are,
\ben \label{TDGLEqns+Langevin}
\ba{rcl}
\tau_0\ddot A_{\al i} 
&+&
\gamma\dot A_{\al i} 
+ 
  \al  A_{\al i} 
 -
 K_1\pa^2 A_{\al i} - (K_2+K_3) \pa_i\pa_j A_{\al j} 
\\
&&+ 2\left[
\be_1 A^*_{\al i} \Tr{A A^T} + 
\be_2 A_{\al i} \Tr{A A^\dagger} \right.
\\
&&\qquad
+ \left.
\be_3 (A A^T A^*)_{\al i} +
\be_4 (A A^\dagger A)_{\al i} + 
\be_5 (A^* A^T A)_{\al i} 
\right] 
= 
\noise_{\alpha i}(\br,t)
\,.
\ea
\een

For temperatures very close to $T_c$, i.e. the ``gapless region'' where $|\Delta(T)|\ll \pi \kb T_c$, the damping by the fermionic bath is given by~\cite{miz23}
\begin{equation}
    \gamma = \hbar \frac{\pi N(0)}{48\kb T_c}
    \,.
\end{equation}
However, $\gamma$ decreases rapidly below $T_c$ as the mean field order parameter, and excitation gap, become established. At low temperatures the temporal dynamics is dominated by the inertial term defined by Eq.~\eqref{eq-TDGL_kinetic_term}. However, at intermediate temperatures we retain both the inertial and damping terms.
On dimensional grounds 
It is convenient to express $\gamma = \hbar^2 \tilde\gamma (\sqrt{35\zeta(3)/60}) 
N(0)/6\pi (\kB\Tc)^2$, where $\tilde\gamma$ has dimensions of frequency.

\subsection{\label{KZVTDGL:lattice}Dynamic lattice field theory simulations}

Simulating order parameters living on a high dimensional manifold in 3+1 dimensions requires high-performance computational resources, both in terms of parallel floating point performance and effective input and output (I/O) for visualisation of the results.  
In the last two decades, high performance computing (HPC) technologies have improved remarkably both in hardware and software branches. 
One of goals of QUEST-DMC project is developing libraries to solve and analyze static and time-dependent GL equations easily, in the way of distributed parallelism and parallel I/O. 
We solve the TDGL equations in Eq.~(\ref{TDGLEqns+Langevin}) with finite difference discretization and explicit time discretization \cite{Strikwerda2004FD}. 
In order to write Eq.~(\ref{TDGLEqns+Langevin}) in dimensionless form, the order parameter $A_{\al i}$ is expressed in units of $\kB \Tc$, the length unit is the zero-temperature limit of the GL coherence length $\xiGL=\sqrt{7\zeta(3)/20}\,\xi_0$, and the time unit is $\tGL = \sqrt{5/3}\,(\xiGL/\vF)$.
Then, the discretized equations are evolved in time for each point on a Cartesian grid in three space dimensions. 
To utilize HPC systems with distributed memory, we use the lattice field theory library HILA as our framework \cite{HILARepo,Laine:2022}. 
HILA offers a uniform grid, on which the number of sites along each Cartesian direction can be chosen separately, and a series of pre-defined class templates to easily handle scalars, vectors and matrices as physical fields on the lattice. 
It also has excellent scaling with number of lattice sites. 

Another significant aspect of simulating dynamics with TDGL theory are the boundary conditions and initial conditions. 
We adopt periodic boundary conditions and surface scattering boundary conditions to model different experimental situations~\cite{ambegaokar1974landau,wim19,wim16}.
When we test scenarios, e.g., Baked Alaska or the Cosmological scenario, which rely on physical processes in the bulk of sample periodic boundary conditions can be used. On the other hand, for heterogeneous environments e.g., order parameter distortion or textural singularities near physical boundaries, a wide range of boundary conditions that take into account different levels of atomic scale roughness have been developed~\cite{AmbegaokardeGennesRainer1974,nag96a,vorontsov2003thermodynamic,vor18}. 

We set the initial conditions for the order parameter configuration on the 3-dimensional  spatial lattice for various homogeneous equilibrium states such as A, B or normal phase. Different levels of noise can be introduced to simulate the fluctuating forces from the thermal excitations. In the rest of Sec.~\ref{KZVTDGL}, we discuss our simulation results with this types of initial states and let those generated from other situations to be discussed in future works.

\subsection{\label{KZVTDGL:results}Preliminary results for a highly disordered initial state}

Here we discuss a test of the Cosmological scenario using the numerical technology which we introduced in Sec.~\ref{KZVTDGL:lattice}.
In order to gain a general understanding of our numerical tool kit, as well as the features of physical system built upon it, we set up a statistically homogeneous noisy initial state, with material parameters, $\alpha(T)$ and $\beta_p(T)$,  with uniform and fixed temperature and pressure. This can be thought of as modelling a 
quench with cooling rate $1/\tau_{Q} \rightarrow \infty$ over the simulation grid with non-equilibrium order parameter configuration \cite{Zurek:1985qw,VolovikCzechJP1996}. 

The spatial grid was $512\times 256\times 256$ sites with a lattice spacing  
$0.5\,\xi_\text{GL}$. 
We chose to simulate dynamics at $p=25$ bar and $T=1.228$ mK, for which 
$\xi_\text{GL} = 12.0 \, \text{nm}$, making the larger side length 6.2 $\mu$m.
The temperature-dependent Ginzburg-Landau coherence length at the given temperature is $\xi_\text{GL}(T) = 17.6 \, \text{nm}$. 
The total simulation time was $10^{3} \tGL$, corresponding to $665$~ns based on $\tGL=0.65$ ns at this pressure and temperature. The damping parameter was set to $\tilde\gamma = 0.02 \tGL^{-1} = 30.05$~MHz. 
This choice is somewhat larger than the value inferred from the absorption of sound near the $J=2^-$ Higgs mode in $^3$He-B~\cite{HalperinPhysica1982,hal90}. The values based on experimental measurements are order $0.1-5$~MHz. 

The spatial boundary conditions were chosen to be periodic in all three directions.
This boundary condition allows us to focus on the possible nucleation of B-phase during a simulation in the absence of confining boundaries.
The initial configuration of $A_{\alpha i}$ was chosen from a Gaussian distribution at every site without spatial correlations, but with mean order parameter corresponding to the equilibrium A phase.

\begin{figure}
    \centering
        \begin{subfigure}{0.9\linewidth}
        \centering
        \includegraphics[width=0.95\linewidth]{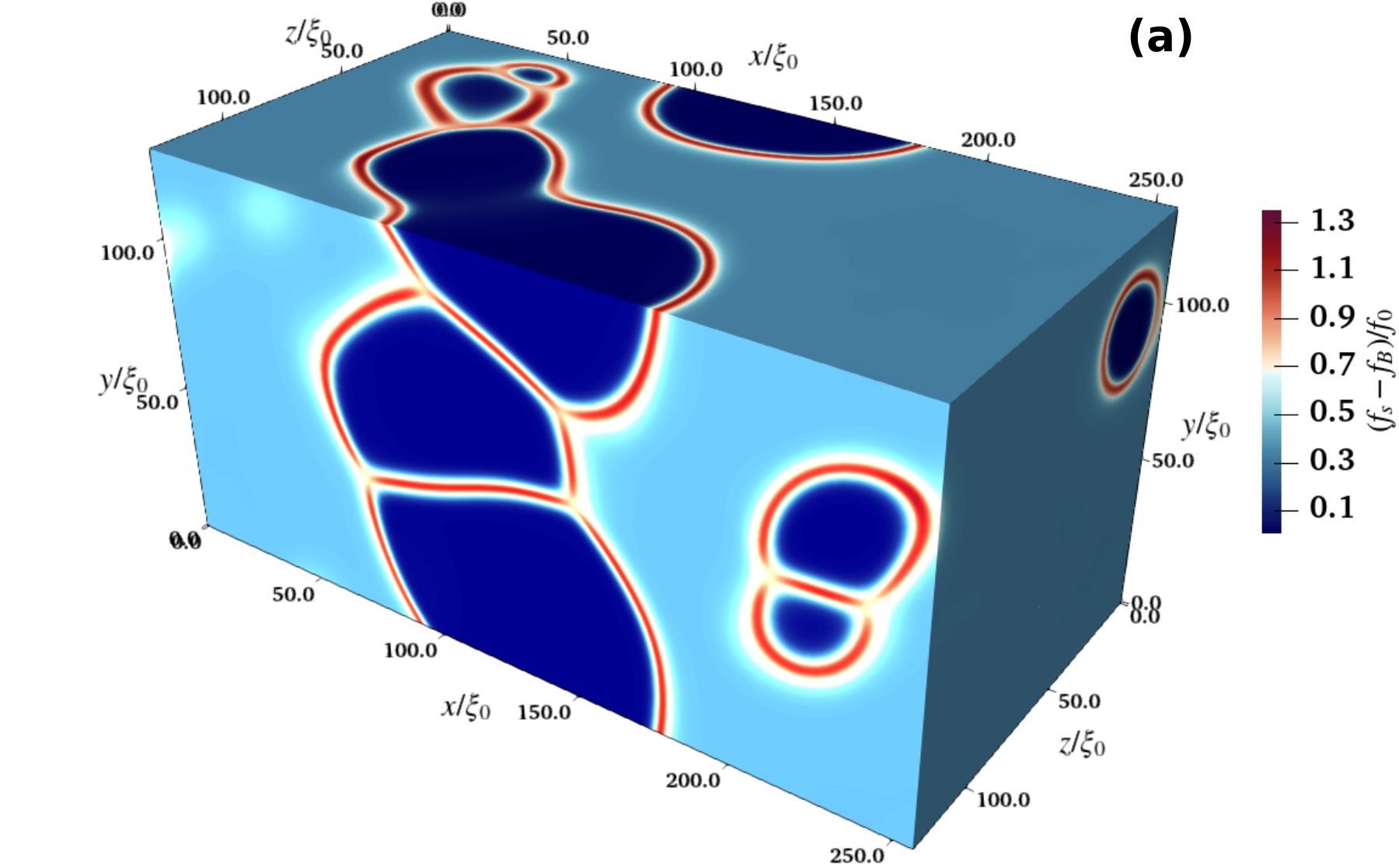}
        \end{subfigure}%

        \begin{subfigure}{0.9\linewidth}
        \centering
        \includegraphics[width=0.95\linewidth]{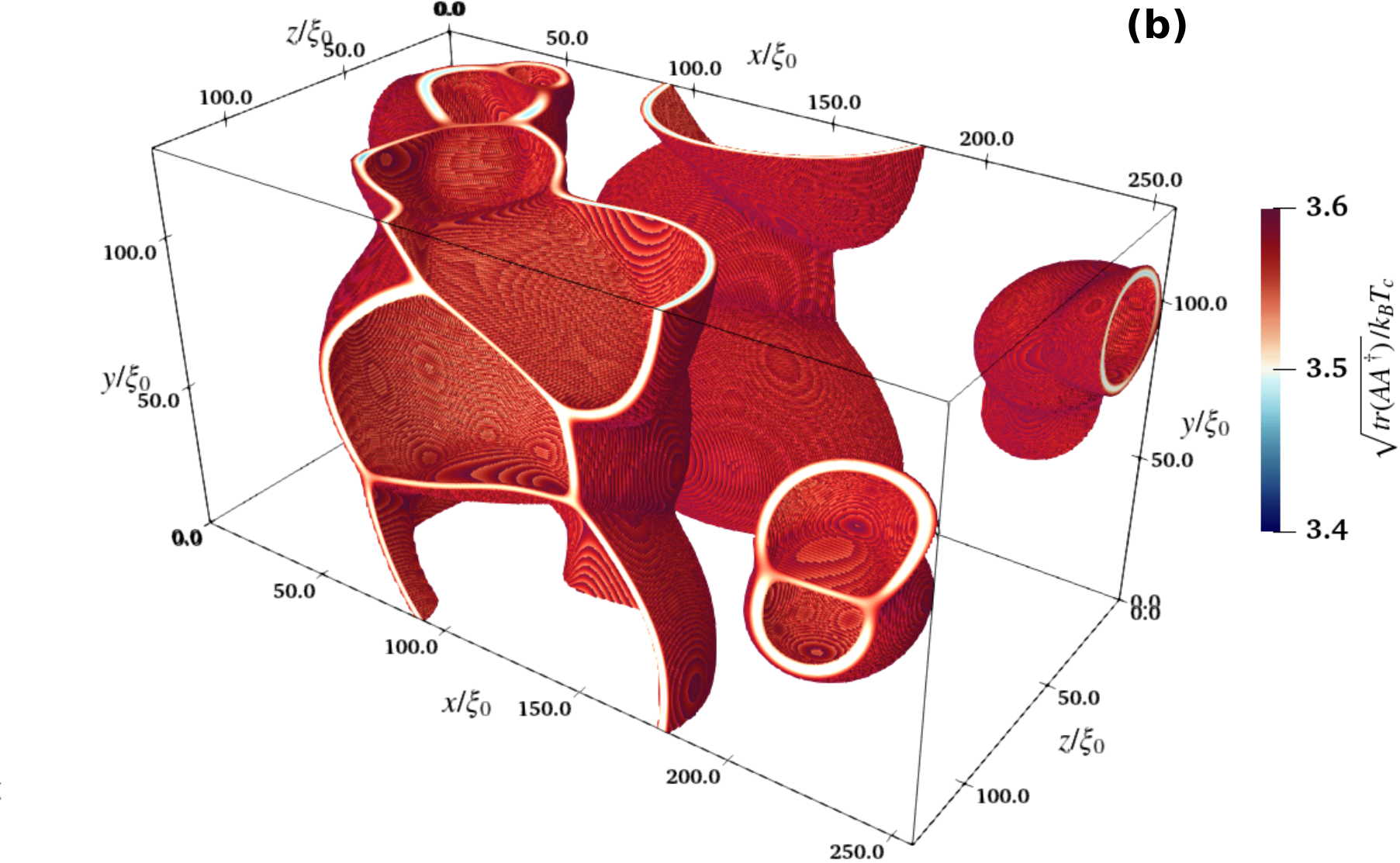}
        \end{subfigure}%

    \caption{Snapshots of distributions of stationary free energy $f_{\text{s}}$ in Eq.~(\ref{eq-fGL}) and $\sqrt{\Tr{AA^{\dagger}}}$ at running time $t=91.7$~ns.
     (a) Distribution of stationary free energy $f_{s}$ over simulation grid as described in Sec.~\ref{KZVTDGL:lattice}. 
     We calibrated $f_{\text{s}}$ against bulk free energy $f_{\text{B}}$ of equilibrium B-phase, then the dark blue corresponds of bulk B-phase, while cyan corresponds to bulk A-phase.
     The domain wall has higher stationary energy because of the gradient energy. 
     (b) Profiles of $\sqrt{\Tr{AA^{\dagger}}}$ between $3.4$ and $3.6$, corresponding to the range found in domain wall structures at the temperature and pressure of the simualtion ($1.228$ mK, $25$ bar). 
     The domain walls visible are either between A- and B-phases or between different B-phases.
     The latter are non-topological domain walls, which require specific symmetry between two domains \cite{SalomaaPRB1988}. 
     The coexistence of A-B domain wall and B-B domain walls can be understood as defects nucleated during fast quench, which is a natural and expected result in cosmological scenario \cite{VolovikCzechJP1996}.
     }   
    \label{fig:ABWallFreeEnergyDensity}  
\end{figure}

With the sufficient onsite noise, which creates a ``glass-like" order parameter, we obtain B-phase bubbles that appear after a run time $t_{n} \simeq 50$~ns, which is significantly longer than the characteristic pair formation time $\hbar/\sqrt{\Tr{AA^{\dagger}}}$ $\sim 1$~ns. 
Interestingly, increasing the onsite noise amplitude does not necessarily lead to an increased likelihood of the nucleation of B-phase. We found that only noise amplitudes in the range $0.4\Delta_A - 0.5\Delta_A$, where $\Delta_A$ is equilibrium A-phase gap, triggered nucleation of B-phase in this particular simulation.
Figure~\ref{fig:ABWallFreeEnergyDensity} shows the B-phase bubble (dark blue) that have nucleated in the surrounding metastable A-phase (cyan), and the corresponding A-B domain walls (red) after time $t=91.7$~ns. Specifically, Fig.~\ref{fig:ABWallFreeEnergyDensity}(a) shows the distribution of locally stationary free energy density $f_\text{s}$ defined in Eq.~(\ref{eq-fGL}), which is calibrated against the magnitude of the bulk free energy density $f_{B}$ of B-phase. The metastable A-phase free energy density is depicted in cyan, while A-B domain walls, which appear in light yellow and red, have much higher energy density contributed in part by gradient energy.

In addition to B-phase bubbles and A-B domain walls, we also find domain walls separating degenerate, but symmetry inequivalent, B-phases, as shown in Fig.~\ref{fig:ABWallFreeEnergyDensity}(a) and (b).
Such walls and their relevance to Cosmology have been first considered in Ref.~\cite{SalomaaPRB1988}. In isotropic pure \hethree\ \cite{mukharsky2004observation} these domain walls are non-topological, but can exist when a certain symmetry is present between two different B-phase domains \cite{SalomaaPRB1988,silveri2014hard}. 
The specific symmetry group involved in this situation consists of $\pi$-rotation of global $U(1)$ phase and global $\pi$-rotations of $SO(3)_{\mathbf{S}}$ in spin space.  
This interesting coexistence of A-B domain walls and B-B domain walls is to be expected in the Cosmological scenario based on the degeneracy space of the bulk B-phase. 

By contrast, symmetry-breaking fields, such as that imposed on \hethree\ by nematic aerogel, lead to topological protection \cite{MakinenZhangEtsovJPCM2023,Zhang2020PRR2020,VolovikZhangPRR2020,Kibble82,Eto23}], analogous to Cosmological Lazarides-Shafi domain walls. In \hethree\ confined to slab geometry, such domain walls are at the heart of the putative crystalline superfluid phase with spontaneously broken translation symmetry \cite{vor07,wim16,levitin2019evidence,yapa2022triangular} (see also \cite{shook2020stabilized,levitin2020comment,shook2020comment} for transport experiments with stepped confinement). 
Here the mechanism of nucleation of the domain walls is a major outstanding question, that may be related to the A-B transition puzzle. In addition to the ``hard'' \cite{silveri2014hard} domain walls discussed above, ``soft'' textural domain walls can be stabilised and manipulated by a combination of confinement, magnetic field and nuclear dipolar energy \cite{levitin2013surface}.

These preliminary studies with a homogeneous quench and infinite cooling rate do not allow us yet to draw firm conclusions about the nucleation of B phase bubbles in metastable A-phase. However, we note that the order parameter remains non-zero everywhere throughout the evolution, suggesting that the bubbles of B phase can appear without the system entering the normal phase following energy injection. 
Simulations based on more realistic temperature profiles, together with physically realistic cooling dynamics, are ongoing and will be discussed in future reports.

\section{Summary and outlook}

In the construction of models of fundamental physics beyond the Standard Model to account for dark matter and the baryon asymmetry, amongst other puzzles, a common prediction is a first order phase transition in the early universe.  Such phase transitions would produce an isotropic stochastic background of gravitational waves.  If the phase transition took place at temperatures at or above the electroweak symmetry breaking scale -- where many models predict new particles and interactions -- the gravitational waves would be potentially observable at future space-based gravitational wave observatories such as LISA \cite{Caprini:2019egz}.   

Computations of the expected signal depend, amongst other things, on a relativistic version of the homogeneous nucleation theory of Cahn and Hilliard \cite{cahn1959free} and Langer \cite{langer1969statistical}. It is therefore important to test the theory in the laboratory. Superfluid \hethree\ has a first order phase transition between the A and B phases, for which the theory predicts that at pressures, temperatures and magnetic fields where the A phase is metastable, the system should remain in the A phase in the course of any conceivable experiment.  Yet the transition usually happens within in a few hours \cite{schiffer1995PLTPnucleation}. The QUEST-DMC collaboration aims to resolve this puzzle, and to decide whether the experimental observations point to a new rapid bulk nucleation mechanism, or to explanations based on external sources of excitation energy such as high energy cosmic ray particles or radioactive decay products from labaoratory materials~\cite{leggett1984nucleation,bunkov1998cosmological}.

The QUEST-DMC collaboration involves new experiments to control and eliminate boundary effects in two ways: nanofabricated cells taking advantage of confinement stabilization of phases, and utilizing magnetic fields to isolate metastable A-phase from experimental boundaries.  Careful choice of nanotechnology 
makes walls atomically smooth, while shaped magnetic field distributions ensure that the A phase is metastable only in a portion of the experiment not in contact with the walls. In this way surface nucleation sites are eliminated.  Future experiments will use the advances in a parallel strand of work on superfluid \hethree\ as a dark matter detector to understand fluxes of energetic particles and to eliminate them by building experimental facilities underground \cite{QUEST-DMC:2023nug,leason2024this}.

At the same time we are building simulation algorithms and numerical codes to investigate the space-time dynamics of the 18-component order parameter of the simulation system, superfluid $^3$He. We can simulate the evolution of this system disturbed by energy injection, or investigate the distortion of the order parameter around boundaries in complex geometries. In our first set of numerical experiments using the new code, we have shown that
regions of B-phase are nucleated following a spatially uniform random disturbance of the metastable A phase. Sufficiently large fluctuations can produce sufficiently large regions of B phase to overcome the surface tension of the phase boundary, and thus grow into a stable B-phase. 
We plan more detailed simulations of existing nucleation scenarios, including the Baked Alaska model and the Cosmological nucleation scenario, where the disturbance in the order parameter is localised.

\paragraph{Acknowledgments.} We thank Grigory E. Volovik and Vladimir Eltsov for comments. 

\paragraph{Author contributions.}  The manuscript was written by MH, JAS and KZ, with contributions from all authors. The simulation code was written by AE-L with contributions from KZ.  Visualisation code was written by KZ.  The parallel framework code was written by KR. The project was supervised by MH, RPH and JS.

\paragraph{Funding sources.} This work was supported by: STFC grants ST/T006773/1, ST/T006749/1, and ST/T00682X/1; Research Council of Finland grants 320123 and 333609; CSC -- IT Centre for Science, Finland; and the Helsinki Institute of Physics. AL-E acknowledges support from Eusko Jaurlaritza (IT1628-22) and by the PID2021-123703NB-C21 grant funded by MCIN/AEI/10.13039/501100011033/ and by ERDF; “A way of making Europe”. JAS acknowledges past support from the US National Science Foundation, Grant DMR-1508730: ``Nonequilibrium States of Topological Quantum Fluids and Unconventional Superconductors'' and current support from the U.S. Department of Energy, Office of Science, National Quantum Information Science Research Centers, Superconducting Quantum Materials and Systems Center (SQMS) under contract number DE-AC02-07CH11359. \

\paragraph{Competing interests.} The authors declare no competing interests.

\bibliography{He3PT,phase_transitions,SFnSC}

\end{document}